\documentclass{llncs}
\usepackage{makeidx}  % allows for indexgeneration
\usepackage[ansinew]{inputenc}
\usepackage{graphicx}
\usepackage{url}
\usepackage[linesnumbered]{algorithm2e}
\usepackage{amsmath}
\urldef{\mailsa}\path|christian.vorhemus@univie.ac.at|
\urldef{\mailsb}\path|erich.schikuta@univie.ac.at|
\newcommand{\keywords}[1]{\par\addvspace\baselineskip
\noindent\keywordname\enspace\ignorespaces#1}

\begin{document}
\frontmatter          % for the preliminaries
\pagestyle{headings}  % switches on printing of running heads
%\addtocmark{Chord} % additional mark in the TOC
%
\title{Blackboard Meets Dijkstra for Optimization of Web Service Workflows}
%
%\titlerunning{Chord}  % abbreviated title (for running head)
\author{Christian Vorhemus \and Erich Schikuta}
\authorrunning{Christian Vorhemus & Erich Schikuta} % abbreviated author list (for running head)
\institute{Faculty of Computer Science\\
University of Vienna\\
\mailsa\\
\mailsb\\
}

\maketitle              % typeset the title of the contribution
\begin{abstract}
This paper presents the integration of Dijkstra's algorithm within a Blackboard framework to optimize the selection of web services from service providers. In addition, methods are presented how dynamic changes during the workflow execution can be handled; specifically, how changes of the service parameters have effects on the system.
For justification of our approach, and to show practical feasibility, a sample implementation is presented.

\end{abstract}
\section{Introduction}

Cloud computing has become a frequently used buzzword in recent times, though in computer science it has been an often discussed topic for years. The main idea behind this notion is providing resources like texts, pictures, videos or even calculation time and disk space via the Web and is, viewed in this light, as old as the Internet itself. However, in the early days of computer-assisted business, the idea of accessing resources via a network was not present, in fact, every enterprise owned locally delimited IT hardware resources to support the own business process. Establishing a working IT was an expensive but necessary capital expenditure (CAPEX).
Nowadays, we see the usage of resources in a different light. Like the advent of object oriented programming changed our sight about programming, the consideration of resources as services will have a similar deep influence for future computer science. Especially in the economic sector, the outsourcing of IT is still on the agenda. In this context, a critical decision has to be made in every larger enterprise: Should the IT be a capital or operational expenditure (OPEX)? Or in other words: Should disk space or calculation time be provided by third parties as services or should all necessary resources be provided by one's own? This "CAPEX vs. OPEX"-question is more and more decided in favor to an outsourced OPEX-solution. This change, however, raises a number of issues. In a business context, one of the most important questions is: How can we get the services with the lowest costs and the largest benefits to do our job? This question also crucial for computational workflows which are cloud sourced.
The goal of this paper is to present a novel approach for the optimization of web service selection under dynamic constraints. Hereby, we propose an artificial intelligence approach by mapping the service selection problem to a graph representation and to apply a combination of Dijkstra's algorithm and Blackboard method for optimization.
%This is the main motivation for this paper and will be discussed within the next sites.
The layout of the paper is as follows: In section~\ref{sec:related_work}, an overview about the related work is given. The Blackboard method, its optimization approach and the specifics of dynamically changing environments is presented in section~\ref{sec:the_blackboard_approach}.
Section~\ref{sec:optimization_process} gives the definition of used optimization algorithms. Finally a sample implementation is presented for justification of our approach.

\section{Related Work}
\label{sec:related_work}

The usage of the Blackboard approach for grid workflow optimization was presented in~\cite{one,two,two.1}. In these papers the problem of changing conditions during the execution of the algorithm was pointed out. In a distributed environment like the Web, the availability of all necessary services to complete a workflow is not guaranteed. Furthermore it is also possible that users change their configuration. To consider changed condition, the authors used the A-* algorithm and extended it by a "blockedlist" which contained not available services. Thanks to the A-* it is not necessary to know all services at the beginning of the execution. In case of the change of a service parameter, services may also transfer back from the closedlist to the openlist, depending on the new, updated costs.
The preceding approach was extended by~\cite{three,three.1,three.2,three.3} with a multistage hierarchical solution. They used the Blackboard framework to optimize query-execution plans in databases. Furthermore, they outsourced a piece of a local optimization process to another source and worked with a hierarchical approach.
A fundamental description of how services can be qualified is presented in~\cite{four}. Among an UML-based approach to classify QoS-Attributes, also a detailed description of how services can be combined is given. Like in different other approaches, the authors also prefer an XML-based language for inter-service-communication.

\section{The Blackboard Approach}
\label{sec:the_blackboard_approach}

\subsection{Blackboard Architecture}

The Blackboard framework consists of four different elements:
\begin{itemize}
	\item A global Blackboard: The Blackboard is a data model consisting of different partial solutions, which are contributed by the brokers. This idea can be depicted as a wall on which different experts collect their knowledge and form the partial solutions to a global optimal solution.
	\item Resources: A resource is, as described in~\cite{two} any kind of service which provides functionality that is needed to finish a subtask of a workflow.
	\item Broker/Agents: Agents are autonomous pieces of software. Their purpose in this architectural framework is finding suitable resources for the Blackboard. An Agent can be a service too and therefore also be provided from external sources.
	\item Controller: There are different kinds of controlling components, which are described section~\ref{sec:optimization_process}. Their main purpose is controlling the start of the algorithm, controlling the brokers and bringing the results back to the user.

\end{itemize}

\begin{figure}
	\centering
		\includegraphics[width=300px]{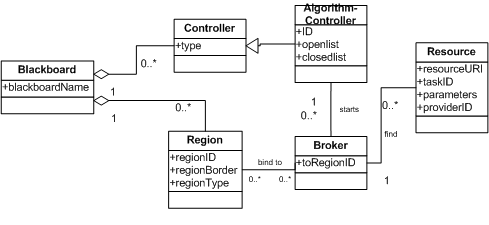}
	\caption{UML diagram of a Blackboard architecture}
	\label{UMLdiagram}
\end{figure}

Figure~\ref{UMLdiagram} shows an UML-based notation of how a Blackboard can be described generically. A Blackboard consists of regions (they refer to subtasks, as described in the next section) and controllers. The algorithm-controller stores an openlist and a closedlist. The openlist contains all promising nodes which have to be analyzed, the closedlist contains all nodes which are finally observed.

Resources are found by Brokers which are bound to specific regions. In this context it also seems necessary to describe, what a resource or a service actually is. A services realizes a generic functionality, offered by a service provider and used by a client. Everything what is needed by a client and provided by a service-provider can be a service: If a client needs disk space, the offer of storing data can be as well a service as the assumption of complex processes. Usually, a single service is not sufficient to fulfill the needs of a user. For example, if a user wants to convert videos and store them, two services are needed: One to convert the videos and one to upload and save them in the cloud. From the users perspective, the two coupled services are considered as one. In our example, the user just passes the video as an input, the first service will convert it and automatically without any further inputs of the user, transmit it to the second service which will finally upload and store it. The user does not care about service-communication or service-composition, he just wants a cheap and quick solution. In the background, an optimal minimum cost path has to be found.

\subsection{Workflow Optimization}
\label{sec:workflow_optimization}

\begin{figure}
	\centering
		\includegraphics[width=300px]{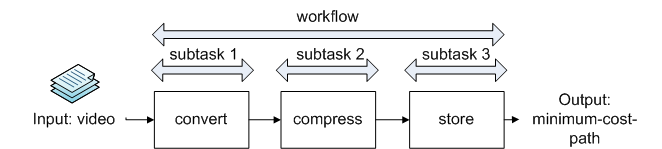}
	\caption{Example of a workflow}
	\label{ExampleWorkflow}
\end{figure}

The Blackboard is divided into several regions and each region represents a subtask of a workflow. Which and how many regions exist depends on the workflow. Let's give a simple example to outline the previous descriptions. Imagine, a user wants to store a video online. He also wants to convert the video from AVI to FLV and compress it to save disk space. The user normally doesn't care, how this workflow is completed in detail, he just wants a good, quick and cheap solution. Furthermore, the user doesn't want to start each subtask manually. He just defines the tasks and hands the video over to the Blackboard, in the end he just gets - in the best case - a message of the successfully converted, compressed and stored video. Figure~\ref{ExampleWorkflow} gives a graphical representation.

The first question which shows up is: What is a good solution for the user? It is necessary to assign a numerical value to each subtask to make a comparison possible. These values are called "costs". Costs are constituted of "Quality of Service" parameters, short QoS. There exist a number of different approaches and formulas like~\cite{five} to combine different notable parameters of a service (like runtime, running costs or success ratio) to a final cost-value. In the next section, a description on how to use the Blackboard approach to find the best provider of a specific service is given.
The second question is: How can a user give restrictions on the workflow? In our example, the user may need a minimum of 15GB of disk space to store the video, so it only makes sense to search for services which provide more than 15 gigabyte space. All these rules are input parameters for the Blackboard. Lets continue with our example: The value for converting the video should be smaller than 60, the compression-ratio has to be greater than 20, the disk space has to be greater than 15, so

\begin{align}
convert \leq 60\\
compress \geq 20\\
store \geq 15
\end{align}
%\vspace*{.4cm}

We base our cost of service calculation on the mathematics of~\cite{one}. Thus, for maximizing respectively minimizing a region the following equations are used Hereby x denotes an "offer" of a service and border(x) the value that is at least acceptable for the execution of the workflow:

\begin{align}
x_{max}& = \frac{border(x)+1}{x+1}
\end{align}

for a minimization we use

\begin{align}
x_{min}& = \frac{x+1}{border(x)+1}
\end{align}

\begin{figure}
	\centering
		\includegraphics[width=300px]{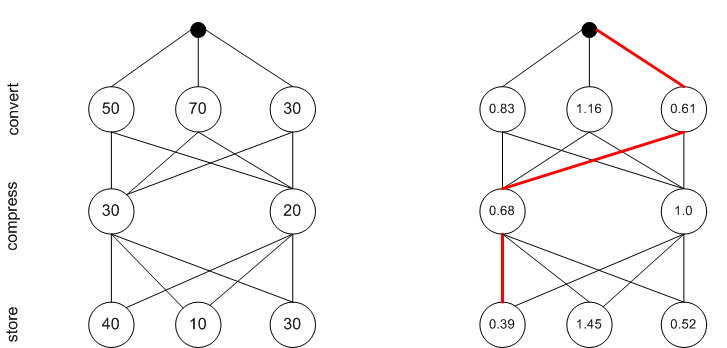}
	\caption{Minimum cost path}
    \label{MinCostPath}
\end{figure}

To underline what is said so far, another example is given. Figure~\ref{MinCostPath} shows a simple Blackboard consisting of 3 regions: convert, compress and store. For region "convert", three different services are available. To all services, a value is assigned. In this example we assume that there is only one cost-parameter for each service, examples for more than one parameter are shown in the next section.

We now have to find services which fulfill our restrictions, in other words, for the first region we search all services with an offer less than 60. We use the previously discussed formulas and calculate the costs for each service, afterwards we just look for the service with the minimal costs of each region. These are our optimal services. Of course, these tasks can be parallelized: It's not necessary to complete the search for an optimal service for region 1 before we can start the search for region 2. We can use this fact to speed up the algorithm.

\subsection{Finding the Best Service Provider}
\label{sec:finding_the_best_service_provider}

\begin{figure}
	\centering
		\includegraphics[width=300px]{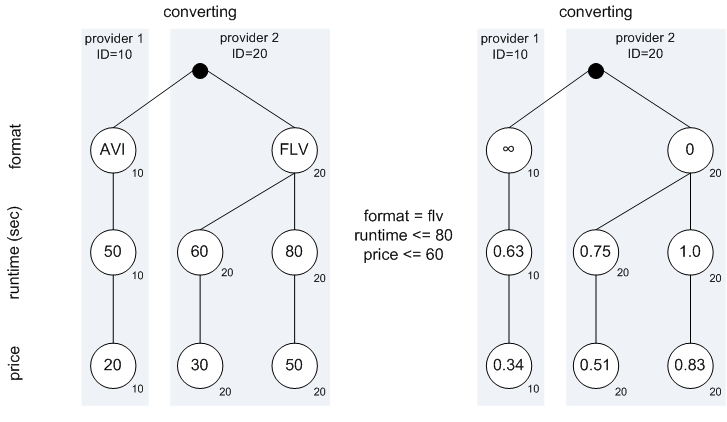}
	\caption{Calculation of the costs of each node}
	\label{ClacCostsEachNode}
\end{figure}

In the majority of cases, there is more than one value to describe the costs of a service. For example, the user is not only interested which services offer a disk space greater than 15GB, but he also includes the price for the storage in his calculations. To handle this scenario, we list all parameters of all providers which come into consideration and connect them, the result is a graph.
Figure~\ref{ClacCostsEachNode} shows the subtask "convert" and two providers (with ID 10 and 20). The first provider offers the conversion to AVI, the second offers the conversion to FLV and gives two more options to choose from, a faster and more expensive possibility (runtime) and a slower option with a lower price. Again, the user sets restrictions; in this example he chose FLV as output format, a runtime less than 80 and a price less than 60.
After we apply the formulas above to our parameters, we get the cost for each node. Note, that in case of Boolean-parameters, the costs are set to zero if the condition is true and set to infinity, if the condition is false.

To find the best provider, we use an algorithm to find the lowest cost path. The Dijkstra algorithm is an appropriate choice for this purpose. In some cases, the estimated costs for a node may be known in advance, for example in databases. Most of the modern database engines use heuristics to assess the estimated costs to speed up the query. If a good Estimator is known and the risk of overestimating the costs is low, we can use the A-* algorithm instead of Dijkstra's algorithm to calculate the total costs.

\subsection{Dynamic Changes}
\label{sec:dynamic_changes}

By now, we just focused on a static environment but changes during the workflow execution are likely to occur when we use web services. Therefore we now take a look on what can happen in dynamic environments.
Dynamic changes can be divided into three groups, namely

\begin{itemize}
	\item Changes of settings by the user,
	\item Changes of services, and
	\item Changes of experts.
\end{itemize}

\begin{figure}
	\centering
		\includegraphics[width=300px]{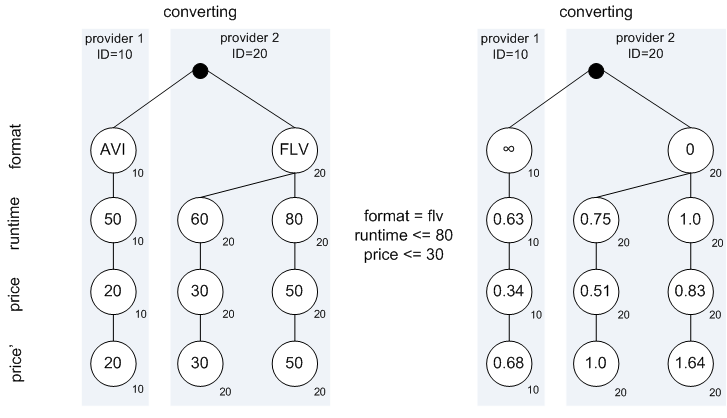}
	\caption{Change of a rule}
	\label{ChangeRule}
\end{figure}

If the user changes the settings, he has three options: He can add a rule, he can modify an existing rule or he can delete a rule. In case of a change, it is unwise to recalculate the whole lowest cost path from scratch, in fact it is not necessary at all. If a rule is added or changed, the new one is just put at the end of the graph. Figure~\ref{ChangeRule} shows an example: Let's assume, the user wants to lower the border of parameter "price" from 60 to 30. We just add a new region price' and calculate the costs of the parameters with our new borders. Of course it is also possible to start a loopback to the initial price-region and recalculate the costs. However this is not necessary and more difficult to implement.

\begin{figure}
	\centering
		\includegraphics[width=300px]{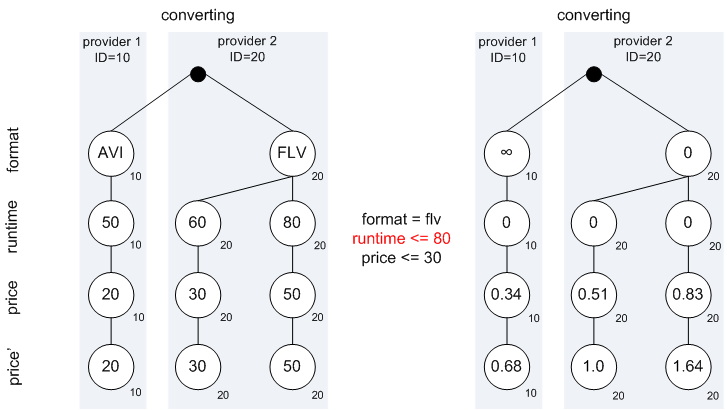}
	\caption{Deleting a rule}
	\label{DeleteRule}
\end{figure}

If a user deletes a rule while the algorithm is running, the costs of all regarding parameters are set to zero. Figure~\ref{DeleteRule} shows this scenario in our example: If a user deletes the runtime-restrictions, the runtime-parameters becomes 0 or, in other words, the values will no longer be considered.

Problems occur, if we have already found a lowest cost path and the user deletes rules afterwards. In this case we cannot just ignore the deleted parameters in our lowest cost path because due to changing parameters, it is not possible to ensure that the lowest cost path found so far has the total minimal cost - maybe a new and better way arises. A possible solution for this problem was given in~\cite{four}. The authors suggested to use orthogonal lists in which the costs of all nodes are saved, though this solution is very memory consuming, if many nodes have to be considered. Another possible way to deal with this problem is a partial recalculation based on a "backtrace mechanism. Since the ancestor of each node is known, we just track our lowest cost path back to the point where the changes occurred and restart the algorithm. In our example, we would go back to runtime-region, throw all nodes of this region away and load the nodes of the subsequent regions in our openlist. Of course, this method is more time consuming if changes are made at the beginning of the algorithm, though in this solution, no additional memory is needed.

\begin{figure}
	\centering
		\includegraphics[width=300px]{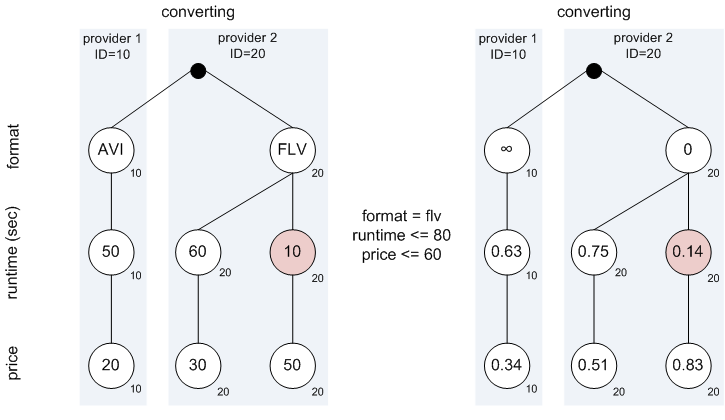}
	\caption{Change of a service value}
	\label{ChangeServiceValue}
\end{figure}

After we described the dynamic changes from users perspective, we will now take a look at changes in services. First of all, the whole service may fail. We use a simple flag called METRIC to indicate whether the service is currently available. If this flag is zero, the service will not be used, 1 means, the service is available. A more difficult question is how changes in parameters can be considered. In each step, the A * algorithm adds the successors of the current node to the openlist and then follows the node with the lowest cost. Nodes which are finally observed are put on the closedlist. A change of cost must be considered in both lists, the openlist and the closedlist. It is also possible that nodes can be transferred back from the closedlist to the openlist.
 Let's look again at an example in figure~\ref{ChangeServiceValue}. Suppose, that the parameter "runtime" will change from 50 to 10. If the concerning node is neither in the openlistlist nor in the closedlist, nothing needs to be done, because the node is not yet known and therefore the costs are irrelevant by now. If the node is already in the openlist, the cost difference between the old and the new value of the cumulative costs has to be calculated.

The result is the new cost of the node. If the node is in the closedlist, the same procedure like in the openlist is done. In addition, the node with the new cost is shifted to the openlist. This is necessary, as by the of the costs of a predecessor node, the costs of the successor nodes are also being changed. The costs of the nodes which succeed the changed one have to be recalculated.

\section{Optimization Process}
\label{sec:optimization_process}

\subsection{Overview}

To describe the architecture of the Blackboard method, the Model-View-Controller concept seems appropriate. All components belong to one of the groups model, view or controller. In the sample implementation a HTML document, which can be opened with a standard Webbrowser represents the "view". The user has the opportunity to set rules via HTML-form, all rules are stored in a shared repository to which only the user and the Blackboard have access. The GUI also displays the results of the algorithm which means - in the successful case - those service providers, which have the best offer. Subsequently, the workflow should be executed autonomously: The user receives a response with the best offer and just has to confirm it. Afterwards, all tasks of the workflow are completed automatically.

\begin{figure}
	\centering
		\includegraphics[width=330px]{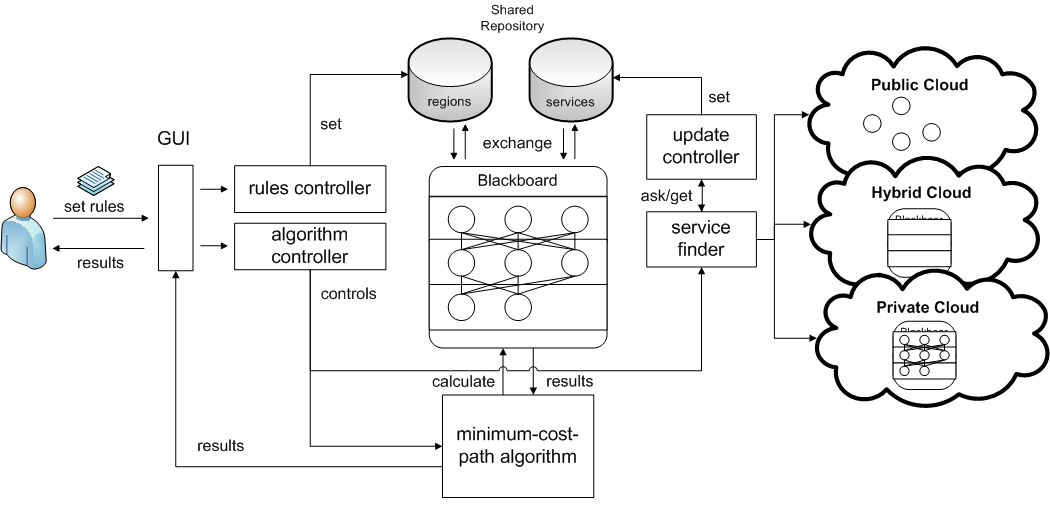}
	\caption{The components of the sample implementation of the Blackboard approach}
	\label{ComponentsImplementation}
\end{figure}

Figure~\ref{ComponentsImplementation} shows the graphical representation of the architecture: The left part shows the layer of the user, the right part is the service-layer. The core consists of the Blackboard, which is implemented in the sample-code as a class and several controllers that govern the process. In addition, a shared repository is used for services and rules as a cache, so that they can be accessed quickly by the Blackboard.

\subsection{Rules Controller and Algorithm Controller}
\label{sec:rules_controller_and_algorithm_controller}

The controller components have administrative functions and are implemented as functions. The "rules controller" controls primarily the settings of rules given by the user. If the user sets constraints for a parameter, for example, "price less than 20" for a subtask, this rule is entered into the shared repository. In addition, this component informs the "minimum cost path algorithm" when a change took place. A change of rules will result - as explained in section~\ref{sec:dynamic_changes} - in a change of regions on the Blackboard. More details on these changes follow in section~\ref{sec:minimum_cost_path_algorithm}.
A central role is played by the "algorithm controller". It initializes the service finder component and monitors the workflow execution. In particular, the algorithm controller delegates subtasks to other components and builds a final solution which will be returned to the user. Algorithm~\ref{algo:1} presents a sample implementation:

\begin{algorithm}[H]
\SetAlgoLined
\SetKwInOut{Input}{input}
\SetKwFunction{executeWorkflow}{Fn}
\Input{List of subtasks in a workflow, artifact (text, video, picture, ...)}
\BlankLine
input = artifact\;
\ForEach{subtask in workflow}{
  bestProvider = findBestProvider(subtask)\;
  output = callServiceProvider(bestProvider, input)\;
  input = output\;
}
return output\;
\SetKwBlock{executeWorkflow}{}{end}
%\SetAlgorithmName{Code}{heuristic}{List of Heuristics}
\caption{Execution of the workflow}
\label{algo:1}
\end{algorithm}
\vspace*{.4cm}

The component receives a list of subtasks. In this context, it must be ensured that each subtask is uniquely identifiable, otherwise it cannot be assigned to a web service. Beside a subtask-list, also an artifact is rendered, for example a video, a text, an image or other media files. The aim is that each subtask is processed on the artifact. Let's look at an example for better understanding: Suppose, the user has an English text as an input. He first wants to translate the text into Spanish, and then he wants this text to be sent as an email to his business partner in Seville. The artifact is the text, that subtasks are the translation and the email. The user additionally gives restrictions, for example, that the translation must be done for free.

In a first step, a function is findBestServiceProvider () is called. This function is used to find die service provider with the minimum total costs (as described in section~\ref{sec:the_blackboard_approach}).
 In a second step, the function callServiceProvider() is called. The function calls a service provider with the artifact (in our example, the English text) and the results (e.g. the translation) are returned. With the result, the next service is called, until the workflow is completed. In a last step, the final solution will be returned to the user.

\subsection{Service Finder and Update Controller}

The Service Finder component is used to communicate with web services. There exists a bunch of different approaches how this communication-process can be performed. In~\cite{six}, the authors argue, that in an RPC-style architecture, every service has its one interface and if a client wants to communicate with it, he needs to know about its semantic. Therefore, they prefer a REST-based approach, especially because of better scalability. However, the communication process is not the main topic of this work, we therefore use just a simple method for client-server communications. To transfer data, the telnet protocol is used, the services are autonomous programs that are located within the SeattleGENI network~\cite{seven}.
SeattleGENI is an open source network testbed similar to PlanetLab. Due to the developers, the Seattle software is currently running on about 1000 machines worldwide. Users can upload their programs written in a constrained programming language called "Repy" (coinage of "restricted" and "python") and running them in a sandbox called "vessel". Each vessel controls the allocation of the hardware resources to the program. In our approach, we configured 10 vessels (10 is the maximal number of concurrently running vessels in the free version) to return a random float number between 0 and 1. This vessel represents a web service, the random number represents an offer of this service.
Since our Blackboard has different regions, we have to find suitable services for each region. So for example, if our workflow consists of converting a video and uploading it, we have to find services which will accomplish the converting part and (in parallel) find services which give us the opportunity to upload it. We assign IDs to each part of the workflow and search for web services with the same ID. Of course, this implicates that our ID is well known by the service. In other words, a standardized directory service like the UDDI would be helpful as a first contact point to find a specific service. However, we assume in our approach that we have a list of services as starting point to solve our task. An example of a possible service description is given below:

[IP=131.12.10.1, PORT=63150, TASK\_ID=25376, METRIC =0, PAR\_LIST = [PRICE, BANDWITH, DISKSIZE], PRO\_ID=10]

A tuple consisting of IP and PORT is used to contact the service (URI), the TASK\_ID is needed to identify which task the service will fulfill, the METRIC represents the availability of the service (0 means, the service is currently not available, a value between 0 and 1 represents the quality of the availability. A very simple way to assign a value for the metric is dividing Round Trip Time (RTT) by 1.), the PAR\_LIST shows, what the service can actually offer (for example disksize, bandwidth or price) and the PRO\_ID identifies the provider of the service. How the data is finally transmitted efficiently (e.g. via a XML-based language or as parameters in a HTTP GET request) is not discussed in detail in this paper.

Closely associated with the service finder component is the update controller component. There are basically two ways how a client can be informed about changes of the web service. First, the web service sends a broadcast to all registered clients (observer pattern). If no request is sent during a specific time interval, the metric of the service is set to 0. Second, the client itself checks if there are any changes of the web service.
How often an update must be performed depends in particular on the following parameters:

\begin{itemize}
	\item the time that has elapsed since the last update,
	\item the variability of the environment, and
    \item the cost of absence of a service.
\end{itemize}

Which type should be preferred depends on the needs of the user. If the user is contractually bound by SLAs for a certain period, it is not necessary to check whether, for example, services at a lower price are available. It is also conceivable, that the service-provider offers discounts to loyal customers; in this case, not the client but the service initialize the contacting.

\subsection{Minimum Cost Path Algorithm}
\label{sec:minimum_cost_path_algorithm}

To calculate the minimum cost path algorithm for our purpose, a modified implementation of Dijkstra's algorithm is defined by algorithm~\ref{algo:2}. The algorithm receives a list of regions (e.g. format, price and runtime) and a list of parameters as an input. Parameters are numerical values, such as price=30. Each parameter is assigned to a service provider (serviceID), for example [price=30, serviceID=10].

The approach is depicted in algorithm~\ref{algo:2}. It starts with the function calculateCosts() to determine the cost of each parameter. Then, the node with the lowest cost is added to the openlist. The openlist contains all know but not yet visited nodes, this process is further described in [2]. After this step, it is checked if the current node is an "endnode" (which means, the last region of the board is reached). If this is not the case, all nodes from the next region, which are provided from the same service provider as the current node are added to the openlist. Finally, the current node is removed from the openlist and added to the closedlist. The closedlist contains all nodes, which are already visited. Then, the procedure is running till there are nodes in the openlist or the exit condition is fulfilled.

\begin{algorithm}
\SetAlgoLined
\SetKwInOut{Input}{input}
\SetKwFunction{retrace}{Fn}
\Input{List of regions, list of parameters}
\BlankLine
openlist = [\,]\;
closedlist = [\,]\;
path = [\,]\;
firstRegion = regionlist[0]\;
lastRegion = regionlist[len(regionlist)-1]\;
\BlankLine
openlist.add(allNodesOf(firstRegion))\;
\BlankLine
\SetKwBlock{retrace}{}{end}
\FuncSty{retrace(Node)}
\retrace{
  path += Node\;
  \eIf{Node.region is firstregion}{
    return}{
    retrace(Node.ancestor)}
}
\BlankLine
\While{openList not emtpy}{
  calculateCosts(openlist)\;
  currentNode = minimum\_cost(openlist)\;
  \BlankLine
  \If{currentNode.region = lastRegion}{
  retrace(current)\;
  }
  \ForEach{parameter in parameterlist}{
  nextRegion = regionlist.index(current.region)+1\;
  \If{parameter.region = nextRegion parameter not in openlist and parameter not in closedlist and parameter.serviceID = current.serviceID}{
      parameter.costs = current.costs + parameter.costs\;
      openlist += parameter\;
      parameter.ancestor = current\;
  }
  }
  openlist = openlist \ensuremath{\backslash} current\;
  closedlist += current\;
}

%\SetAlgorithmName{Pseudocode}{heuristic}{List of Heuristics}
\caption{Find the best provider for a service}
\label{algo:2}
\end{algorithm}
\vspace*{.4cm}

\begin{figure}
	\centering
		\includegraphics[width=300px]{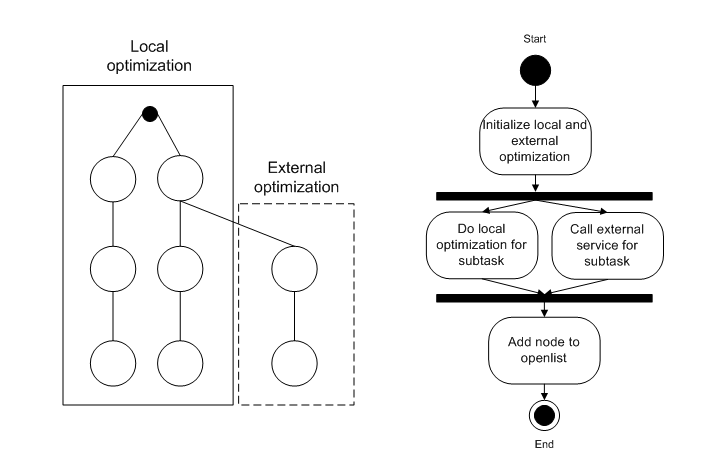}
	\caption{External Optimization}
    \label{fig:ExtOpt}
\end{figure}

By now, it is assumed that the entire optimization process takes place locally. However, it is also possible to outsource subtasks. This is useful, if the optimization problem can be solved more efficiently by external components. However, if the external source can calculate a subtask faster than the local source, can only be assumed in advance. Therefore it makes sense - in analogy to database systems - to start the local optimization process while calling the external source and starting the external optimization process too [3]. If the local source calculates the solution faster, the results of the external source is ignored. Figure~\ref{fig:ExtOpt} gives an overview of the problem: In this example, a branch of the optimization problem is calculated by an external source. If this source returns a solution, it will be added to the openlist by the minimum-cost path algorithm. There are no other modifications of the algorithm necessary, because if the external solution is better than local solutions, it will be added sooner or later to the optimal path (due to less cumulative costs). It is explored by the service-finder component if external optimization is available. If so, this option is marked in the shared repository for the respective regions.

This procedure can be easily included in a hierarchical Blackboard environment: Partial solutions found by one Blackboard can be used as an input for another Blackboard. However, as said before, no modifications have to be made on algorithm, because it doesn't mind where to partial solutions come from, even if there were multiple blackboards involved to get the final solution. The only difference in a hierarchical Blackboard environment is, that a whole Blackboard is treated as a service for another Blackbord.

\section{Working example}
\label{sec:working_example}
\begin{figure}
	\centering
		\includegraphics[width=300px]{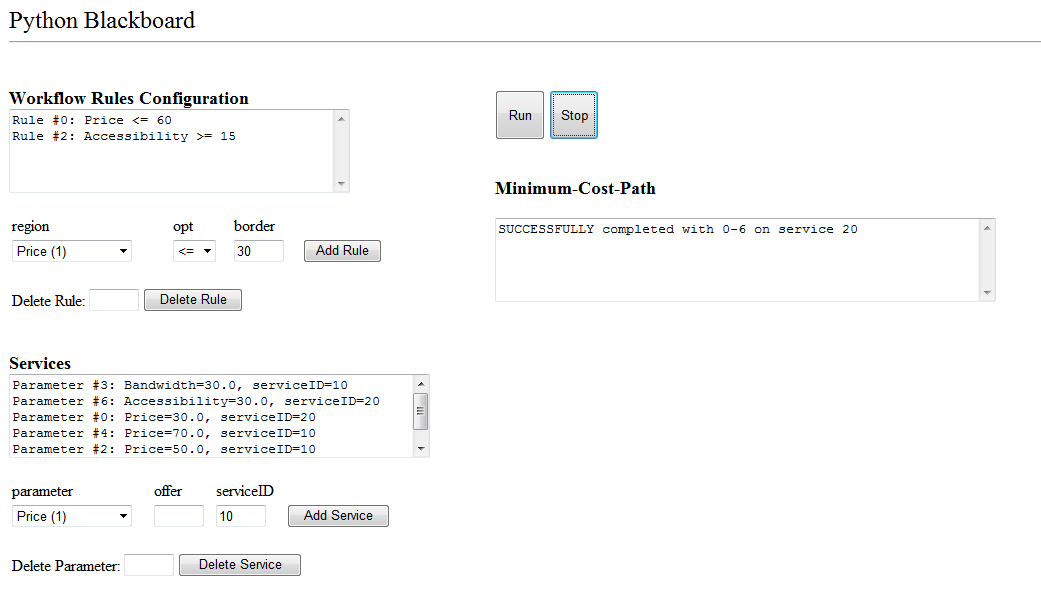}
	\caption{Screenshot GUI Blackboard Application}
	\label{Screenshot}
\end{figure}

For justification the algorithm was implemented within the Google App Engine (GAE)~\cite{eight}. The Blackboard-Application can be found at~\cite{nine}, the architecture of the software follows the concept shown in figure~\ref{ComponentsImplementation}.
The GUI (see figure~\ref{Screenshot}) contains a textarea in the left corner where the rules are defined to describe the workflow. Right below, another textarea can be found where all available services are listed.
In the text area in the right corner the results are displayed. A result consists of the serviceID with the minimum costs and its corresponding parameters (such as price, disk space or bandwidth). If the setting is changed during the workflow execution, the algorithm automatically detects the changes and recalculates the minimum-cost-path.
Measuring call- and response-time on a realistic environment with large datasets is the next step to evaluate our approach.

The picture shows a text area in the left corner. It contains the rules, the user defines to describe the workflow. Right below, another textarea can be found where all available services are listed. In a realistic usecase, however, such a list will not be presented to the user because it is very likely that there are thousands of candidate services and it is the job of the algorithm to search the best ones. In this testing scenario, the service-list is present to check, if the algorithm is working correctly.
In the text area in thew right corner the results are displayed. A result consists of the serviceID with the minimum cost and its corresponding parameters (such as price, disk space or bandwidth). A click on "Run" starts the algorithm and refreshes the results all 5 seconds. If the setting is changed during the workflow execution, the algorithm automatically detects the changes and recalculates the minimum-cost-path.
In our test-scenario, the services are just objects of a service-class but it would be easy to adopt the code to call external sources. SeattleGENI-vessels can be used to simulate working services, spread on servers all over the world. Unfortunately, due to security restrictions, it is not allowed to open sockets within the Google App Engine, therefore it is not possible to contact the external SeattleGENI-"services" in the web engine. Sure enough, such a scenario can be performed local. Measuring call- and response-time on a realistic environment with a large dataset to check, if users can work efficiently with our approach, is the next step to evaluate the approach.

\section{Conclusion and Future Work}
\label{sec:conclusion}
This paper gives an overview about how complex problems can be solved by dividing them into subtasks and search for web services with the best offer for the user. Due to the dynamic environment, it is necessary to consider dynamic changes of services. These changes make partial recalculations of optimal solutions necessary. However, it is a challenging task to limit these recalculations to a minimum. Therefore, suggestions are given, how to deal with these incidental events.

The focus of this paper is on the description of the architectural framework of the Blackboard approach and its components. Suggestions were made, based on the ideas of~\cite{three}, how single optimization-processes can be outsourced to external parts. An important cognition of this analysis is, that for external-optimization, not many modifications have to be made. This leads to the conclusion that it is possible to split the entire Blackboard, outsource all branches and just collect and evaluate the partial solutions locally.

In the future, it should be possible for the user to specify a complete workflow and the computer will automatically find the subtasks which are contained. For example, it should be possible to filter the subtasks from the phrase "Check, if I can play tennis tomorrow afternoon at 3pm in Santa Barbara". In this example, a web service that checks if the tennis court is free or reserved should be called, a web service can check the weather and a web service checks if the user should drive away from home earlier because a lot of traffic is expected.

Summing up, we presented and implemented a framework for optimization of the service selection problem of workflows. The novelty of our approach is mapping the service selection problem to a graph representation and to apply a combination of Dijkstra's algorithm and Blackboard method for optimization. The practical feasibility is shown by a use case implementation.


\end{document}